\documentclass[12pt]{article}
\usepackage{epsfig}
\usepackage{graphicx}
\usepackage{cite}
\usepackage{hyperref}
\usepackage[belowskip=0pt,aboveskip=0pt]{caption}

\def \babar{B{\sc a}B{\sc ar}~}
\def \bea{\begin{eqnarray}}
\def \eea{\end{eqnarray}}
\def \beq{\begin{equation}}
\def \eeq{\end{equation}}
\def \nn{\nonumber}
\def \ok{\overline{K}^0}
\def \({\left(}
\def \){\right)}
\def \s{\sqrt{2}}
\def \ca{{\cal A}}
\def \sx{\sqrt{6}}
\def \osx{\frac{1}{\sx}}
\def \confproc{Proceedings of CKM 2012, the 7th International Workshop on the
CKM Unitarity Triangle, University of Cincinnati, USA, 28 September - 2 October 2012}
\def\Title#1{\begin{center} {\Large {\bf #1} } \end{center}}

\begin{document}
\rightline{UdeM-GPP-TH-12-217}
\rightline{arXiv:1212.NNNN}
\rightline{December 2012}

\Title{Extraction of $\gamma$ from three-body $B$ decays\footnote{\confproc}}

\bigskip


\begin{raggedright}

{\it Bhubanjyoti Bhattacharya \footnote{Speaker} 
and David London \\
Physique des Particules, Universit\'e
de Montr\'eal,\\
\it C.P. 6128, succ.\ centre-ville, Montr\'eal, QC,
Canada H3C 3J7}

\bigskip

{\it Maxime Imbeault \\
D\'epartement de physique, C\'egep de Saint-Laurent,\\
625, avenue Sainte-Croix, Montr\'eal, QC, Canada H4L 3X7}
\bigskip
\end{raggedright}

\begin{quote}
The conventional use of two-body $B$ decays to extract $\gamma$, although theoretically
clean, is currently statistics-limited. On the other hand, a bulk of data on three-body
$B$ decays is available from $B$ factories. Applying the flavor-SU(3)-symmetric approach
proposed in Ref.\ \cite{ReyLeLorier:2011ww} to \babar data, we find the highly promising
result $\gamma = (81^{+4}_{-5} ({\rm avg.}) \pm 5 ({\rm std.\ dev.}))^\circ$. This
establishes the use of three-body $B$ decays as a viable alternative for the extraction
of weak phases. In this preliminary analysis we have neglected several sources of
uncertainties such as the effect of flavor-SU(3) breaking due to meson masses, and error
correlations between input experimental parameters. A better understanding of these will
improve the viability of this method.
\end{quote}

The Cabibbo-Kobayashi-Maskawa (CKM) phase $\gamma$ is conventionally extracted
using two-body $B$ decays \cite{Gronau:1990ka} where the goal is to use the
interference between the decay amplitudes for $b\to c\bar u s$ and $b\to u\bar c s$.
This technique is theoretically clean because of the absence of penguins. However,
current statistics on Cabibbo-suppressed modes of two-body $B$ decays limits the
precision of $\gamma$ obtained in two-body $B$ decays. On the other hand, there is a
fair amount of data available on $B$ decays to charmless three-body final states from
Belle and \babar. However, extraction of weak phases from three-body decays is difficult
because of two reasons. First, the charmless three-body final states do not involve
distinct quark flavors: one has to deal with penguin diagrams. Secondly, even flavor-%
neutral three-body final states are not CP eigenstates : absence of indirect CP asymmetry
makes it harder to extract the weak phase. It was shown in Ref.\ \cite{ReyLeLorier:2011ww}
that in the limit of flavor-SU(3) symmetry these difficulties can be overcome by combining
Dalitz analyses of multiple three-body $B$ decays and considering the part of the amplitudes
that are fully symmetric under the exchange of any two final-state mesons. The purpose
of this talk is to show an application of the method proposed in Ref.\
\cite{ReyLeLorier:2011ww} using experimental data from \babar. The theory details
including the extraction technique are available in Ref.\ \cite{ReyLeLorier:2011ww,
London:2012}.

We have used Dalitz analyses for the following processes available from \babar:
$B^0\to K_S\pi^+\pi^-$\cite{Aubert:2009me}, $K_S K^+ K^-$\cite{Lees:2012kxa},
$K^+\pi^-\pi^0$\cite{BABAR:2011ae}, and $3 K_S$ \cite{Lees:2011nf}. In addition it
is necessary to include data from $B^+\to K^+\pi^+\pi^-$ to estimate SU(3) breaking
as described in Ref.\ \cite{ReyLeLorier:2011ww}. However, our preliminary goal is to
establish the viability of the method and we have ignored effects of SU(3) breaking.
In order to relate the amplitudes of the different three-body decay modes they are
written in terms of flavor-SU(3) diagrammatics. In three-body $B$ decays, neglecting
power-suppressed contributions from weak-annihilation topologies, there are ten
distinct flavor-SU(3) topologies, ``Color-favored Tree'' ($T_1, T_2$), ``Color-%
suppressed Tree'' ($C_1, C_2$), ``Color-favored electroweak Penguin'' ($P_{EW1}$,
$P_{EW2}$), ``Color-suppressed electroweak Penguin'' ($P^C_{EW1}$, $P^C_{EW2}$)
and ``Gluonic Penguins'' ($P_{tc}, P_{uc}$).

The amplitudes for the relevant $B$ decay processes, under the assumption of flavor-SU(3)
symmetry may be expressed in terms of these twelve flavor-topology parameters. In order
to construct a CP eigenstate from the three-body final state, Ref.\ \cite{ReyLeLorier:2011ww}
considers the part of the amplitude that is totally symmetric under the interchange of
any two final state mesons. An added advantage of using the totally-symmetric final-state
amplitude is that the ``electroweak Penguin'' topologies can now be related to the ``Tree''
topologies \cite{London:2012}.

Although it is possible to measure both the direct and indirect CP asymmetries in $B^0\to 3
K_S$, these observables are currently limited by the available statistics \cite{Lees:2011nf}.
This can be implemented by setting $P_{uc}$ to zero. The three-body amplitudes may now be
written in terms of ``effective diagrams'' (linear combinations of the flavor-topology
amplitudes) as follows:
\bea
A(B^0\to K^0K^0\ok)_{\rm sym} &=& a~, \nn \\
\s A(B^0\to K^+K^0K^-)_{\rm sym} &=& - c e^{i\gamma} - a + \kappa b~, \nn \\
2 A(B^0\to K^+\pi^0\pi^-)_{\rm sym} &=& b e^{i\gamma} - \kappa c~, \nn \\
\s A(B^0\to K^0\pi^+\pi^-)_{\rm sym} &=& - d e^{i\gamma} - a +\kappa d~,
\eea
where in terms of the flavor-topology amplitudes $a$, $b$, $c$ and $d$ are:
\bea
a = - P_{tc} + \kappa\(\frac{2}{3}T_1 + \frac{1}{3}C_1 + \frac{1}{3}C_2\)~,
~~~~~~~~~~ \nn \\
b = T_1 + C_2~, ~~ c = T_2 + C_1~, ~~ d = T_1 + C_1~,~~\kappa\sim0.5~.
\eea

The fully-symmetric three-body amplitudes are dependent on the kinematics of the
final-state particles. This dependence can be conveniently represented on a Dalitz
plot. Dalitz analyses for the relevant three-body decays have been performed by
\babar. A convenient model used to represent the dynamics of the three-body final
state is the Isobar model where the final state amplitude is written as follows:
\beq
\ca_{\rm DP} = {\cal N}_{\rm DP}\sum_j c_j e^{i \theta_j} F_j (s_{12}, s_{13})~,
\eeq
where the index $j$ represents an intermediate state, ${\cal N}_{\rm DP}$ serves
the purpose of normalizing the amplitudes to the measured rates for the relevant
processes, $s_{ik}$ is the invariant mass of two final state mesons labeled $i$
and $k$ respectively, and $c_j$ and $\theta_j$ represent the amplitude and phase
of the $j$th isobar coefficient. Although there are three choices for $s_{ik}$,
the sum of all three is a constant and hence only two of them are independent.

The values of $c_j$ and $\theta_j$ were taken from the respective \babar
publications. Although, the error bars on these quantities are correlated,
such correlations are often not available in the literature. In our preliminary
analysis we have chosen to ignore correlations, noting that a more complete
work would necessarily include them. We are now able to compute the totally
symmetric parts of the amplitudes as follows:
\bea
\ca_{\rm sym} = \osx (\ca(s_{12}, s_{13}) + \ca(s_{13}, s_{12}) + \ca(s_{12},
s_{23}) \nn \\
 + \ca(s_{23}, s_{12}) + \ca(s_{23}, s_{13}) + \ca(s_{13}, s_{23}))~.
\eea
A similar exercise can be done to obtain the symmetric part of the amplitude
for CP-conjugate decay process. These may then be used to construct the
observable quantities for every point on a given Dalitz plot as follows:
\bea
X_{\rm DP} &=& |\ca_{\rm sym}|^2 + |\overline{\ca}_{\rm sym}|^2 \nn \\
Y_{\rm DP} &=& |\ca_{\rm sym}|^2 - |\overline{\ca}_{\rm sym}|^2 \nn \\
Z_{\rm DP} &=& {\rm Im}\(\ca^*_{\rm sym}\overline{\ca}_{\rm sym}\)
\eea

A simple count of parameters and observables tells us that one can extract
$\gamma$ in this scenario. For every point on the Dalitz plot we have eight
independent unknown parameters : the magnitudes and relative phases of $a$, $b$,
$c$ and $d$ (one overall phase is arbitrary) and $\gamma$. However, we have nine
observables : $X_{\rm DP}$ from all four processes, $Y_{DP}$ from all but $B\to 3
K_S$ (due to low statistics CP-asymmetry measurements aren't available for this
decay), $Z_{DP}$ from $B^0\to K_S(\pi^+\pi^-, K^+K^-)$. Note that the $K^+\pi^-
\pi^0$ final state is not flavor neutral, and doesn't give us a $Z_{DP}$. The
advantage of this method is that we are able to determine $\gamma$ independently
for every point on the Dalitz plot. Therefore an average over Dalitz plot points
is expected to significantly reduce the error in $\gamma$.

\begin{figure}[htb]
\begin{center}
             \includegraphics[width=0.5\textwidth]{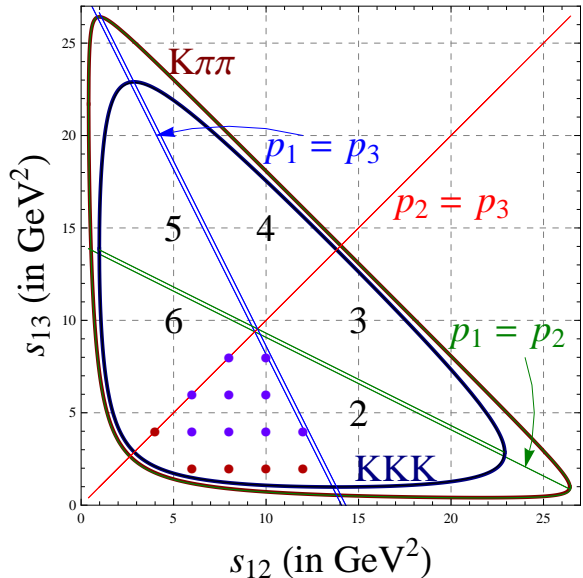}
\caption{Dalitz boundaries and symmetry axes for the four relevant $B$ decays. The
blue dots represent the points that were used for extraction of $\gamma$, while the
red dots represent the points close to the boundary of the Dalitz plots that were
discarded.}
\label{fig:DP}
\end{center}
\end{figure}

In Fig.\ \ref{fig:DP} we show the kinematic boundaries of the Dalitz plots for
the four relevant decay processes. In addition we show the relevant axes of
symmetry, which divide each Dalitz plot into six zones. $\ca_{\rm sym}$ is
totally symmetric under the exchange of any two $s_{ik}$'s. Therefore to avoid
overcounting we are restricted to one of the six zones. Within the chosen sixth
of the Dalitz plots we pick fourteen points to use for the determination of
$\gamma$, which have been displayed in Fig.\ \ref{fig:DP}. We have chosen the
points to lie on a grid with the squared momenta separated by $2 GeV^2$ on either
directions. This choice is completely arbitrary. For each point we perform a
$\chi^2$ fit.
\begin{figure}[htb]
\begin{center}
             \includegraphics[width=0.45\textwidth]{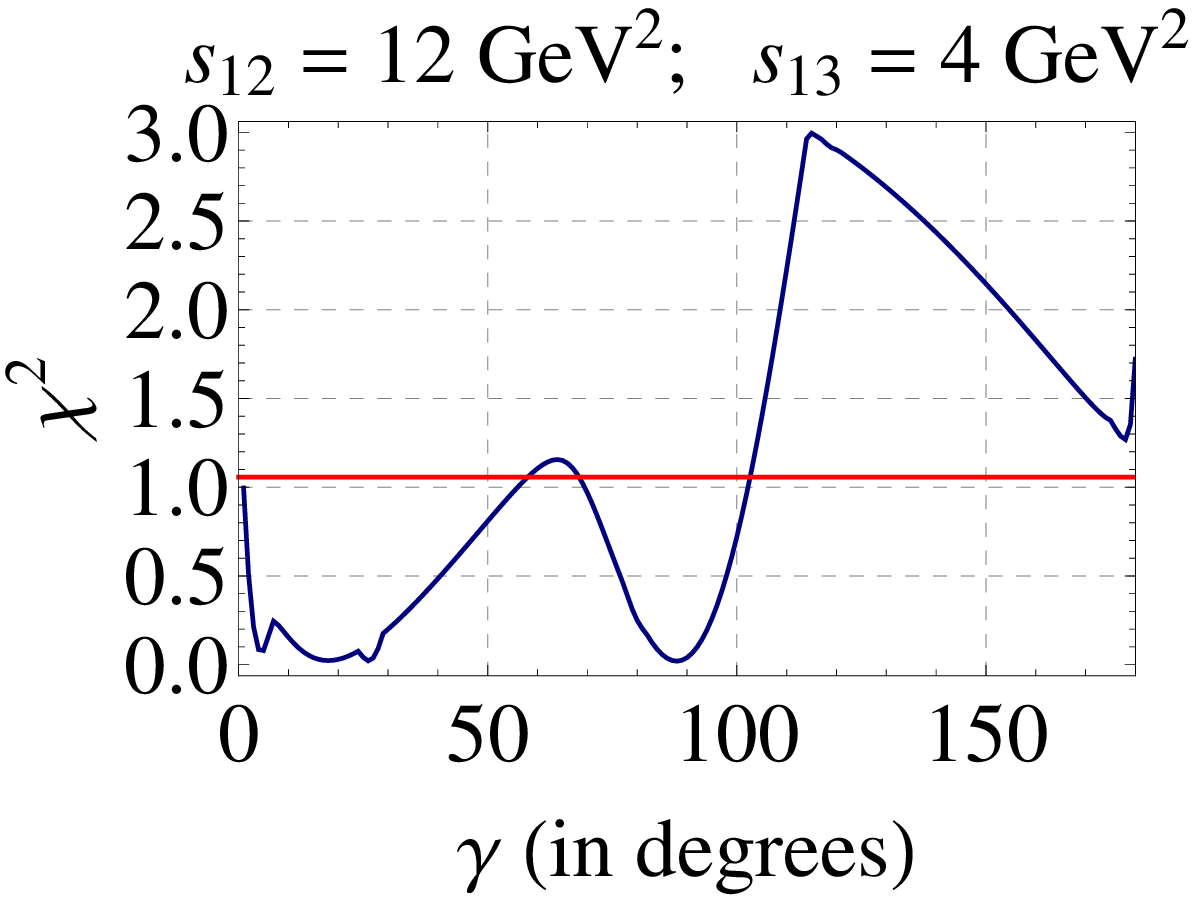} \hskip 0.2in
             \includegraphics[width=0.45\textwidth]{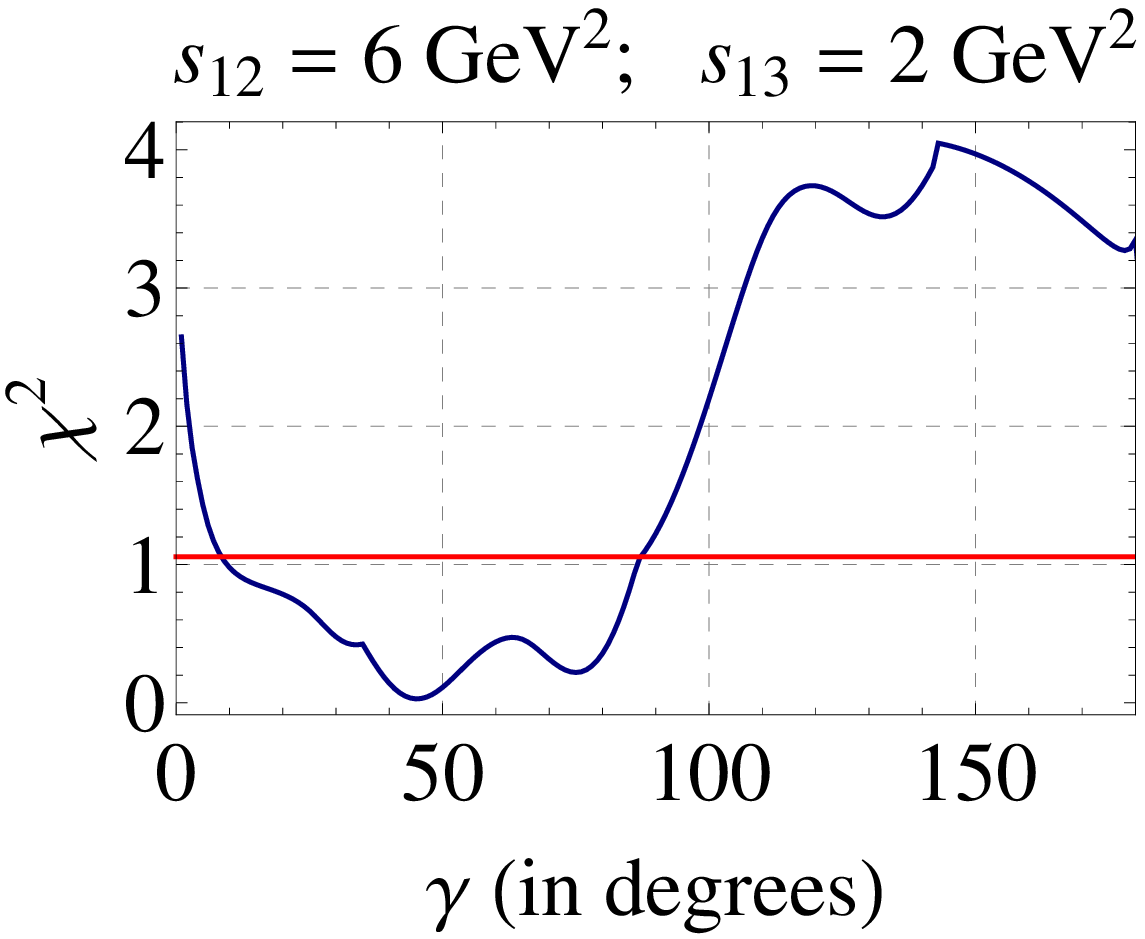}
\caption{$\chi^2$ minimum as a function of $\gamma$ for two different Dalitz plot points.
The plot in the left panel shows one clearly distinguishable solution for $\gamma\sim80^
\circ$. The plot on right shows multiple indistinguishable solutions.}
\label{fig:chi2}
\end{center}
\end{figure}

In Fig.\ \ref{fig:chi2}, we plot the $\chi^2$ minimum as a function of $\gamma$
between 0 and 180$^\circ$, for two different points on the Dalitz plots. The first
plot shows one solution with $\gamma\sim90^\circ$ which can be clearly distinguished
from the other solutions at the $\Delta \chi^2 = 1$ level, and several other
indistinguishable solutions closer to $\gamma\sim30^\circ$. The second plot
corresponds to a point close to the Dalitz boundaries and shows multiple
indistinguishable $\chi^2$ minima. We find similar results for other points close to
the Dalitz boundaries. These solutions require a better understanding of the underlying
dynamics and have currently been left out while computing the final result for $\gamma$.
We find that a clearly distinguishable solution exists for nine points that are further
away from the Dalitz boundaries.

In Table \ref{tab:res} we present our results for $\gamma$ obtained from each Dalitz
plot point where we found a distinguishable solution. A simple mean over the values
of $\gamma$ quoted in Table \ref{tab:res} gives us the following result:
\beq
\gamma = \(81^{+4}_{-5} ({\rm avg.}) \pm 5({\rm std.\ dev.}) \)^\circ~,
\eeq
where the first error was obtained by square averaging over the individual errors and
the second error denotes the standard deviation from the mean of the central values
of $\gamma$ for the nine Dalitz plot points used.
\begin{table}[b]
\begin{center}
\begin{tabular}{|c|c|c|c|}
\hline \hline
$(s_{12}, s_{13})$ & $\gamma$ &$(s_{12}, s_{13})$ & $\gamma$ \\
(in $GeV^2$) & (in deg) & (in $GeV^2$) & (in deg) \\ \hline
 (4, 4) & -- & (8, 8)  & $80^{+13}_{-14}$ \\
 (6, 2) & -- & (10, 2) & -- \\
 (6, 4) & $73^{+12}_{-18}$ & (10, 4) & $86^{+12}_{-14}$ \\
 (6, 6) & $80^{+10}_{-15}$ & (10, 6) & $79^{+12}_{-15}$ \\
 (8, 2) & -- & (10, 8) & $81^{+10}_{-13}$ \\
 (8, 4) & $80^{+12}_{-15}$ & (12, 2) & -- \\
 (8, 6) & $78^{+13}_{-14}$ & (12, 4) & $88^{+14}_{-19}$ \\ \hline \hline
\end{tabular}
\caption{$\gamma$ extracted from several points within the Dalitz plots.}
\label{tab:res}
\end{center}
\end{table}

The error bar on $\gamma$ obtained above is significantly smaller than that
currently obtainable in two-body decays. However, we have neglected a few
sources of errors. For example we have ignored correlations between isobar
coefficients. In practice such correlations may introduce additional error.
Furthermore we have assumed flavor-SU(3) symmetry. However, the masses of the
daughter pions and kaons as well as the intermediate resonances break flavor%
-SU(3) symmetry. A direct consequence of flavor-SU(3) breaking is that the
Dalitz boundaries for a final state $K\pi\pi$ is different from that for $KKK$.
As a consequence points very close to the Dalitz boundaries don't yield
distinguishable solutions for $\gamma$. One way of alleviating this problem is
by including many more points on the Dalitz plot for evaluating $\gamma$. Note
that a more complete analysis of the combined data set for the four Dalitz plots
can avoid errors due to the neglect of correlations between isobar coefficients.
In order to get an idea of how large the effect of SU(3) breaking is, it would
be interesting to include effects of SU(3) breaking due to one extra parameter.

The work presented in this talk was aimed at motivating interest toward
a new method for extracting $\gamma$ from three-body $B$ decays. In this
preliminary analysis we have neglected several sources of errors, however, the
value of $\gamma$ obtained is quite promising. The hope is to find an
alternative viable method for extraction of $\gamma$ using the already
available statistics on three-body $B$ decays from B-factories.

\bigskip
This work was financially supported by the NSERC of Canada (BB, DL) and FQRNT
of Qu\'ebec (MI). BB would like to thank the conference organizers for a wonderful
stay in Cincinnati.

%
%
%

\end{document}